\documentclass[letter]{aa}
\usepackage{graphicx}
\usepackage[varg]{txfonts}

\begin{document}

\title{The Djorgovski-Gurzadyan dark energy integral equation and the Hubble diagram}
\author{H.G. Khachatryan\inst{1,2}, A. Stepanian\inst{1}}
\institute{Center for Cosmology and Astrophysics, Alikhanian National Laboratory, Yerevan, Armenia \and
           Yerevan State University, Yerevan, Armenia}
\offprints{H.G. Khachatryan, \email{harut@yerphi.am}}
\date{Received: 24 August 2020; Accepted: 15 September 2020}

\abstract{We consider the observational aspects of the value of dark energy density from quantum vacuum fluctuations based initially on the Gurzadyan-Xue model. 
We reduce the Djorgovski-Gurzadyan integral equation to a differential equation for the co-moving horizon and then, by means of the obtained explicit form for the 
luminosity distance, we construct the Hubble diagram for two classes of observational samples. For supernova and gamma-ray burst data we show that this approach 
provides viable predictions for distances up to $z \simeq 9$, quantitatively at least as good as those provided by  the lambda cold dark matter ($\Lambda$CDM) model. 
The Hubble parameter dependence $H(z)$ of the two models also reveals mutual crossing at $z=0.4018$, the interpretation of which is less evident.}

\authorrunning{H.G. Khachatryan, A. Stepanian}
\titlerunning{The Djorgovski-Gurzadyan dark energy integral equation}
\maketitle

\section{Introduction}

In his pioneering work \cite{Zel} assigned the cosmological constant $\Lambda$, if non-zero, as a contribution of quantum vacuum fluctuations with the $\epsilon=-p$ 
equation of state  (EOS; energy $\epsilon$ and pressure $p$). After the discovery of the accelerated expansion of the Universe a number of models and approaches were 
proposed (see \citealt{Copeland}), typically involving either theoretical or phenomenological input parameters.

Among the proposed approaches, Gurzadyan-Xue (GX)  models do not have any free or empirical parameters, and they predict a value of $\Lambda$ fitting its observed value. 
This class of models (see \citealt{Khach07a,Khach07,Ver06,Ver06a,Ver06b,Ver07}) also covers those parameters with a variation of fundamental constants, and has been tested 
regarding the Hubble diagram using the supernova (SN) and gamma-ray burst (GRB) observational data \cite{Mosquera}.

Djorgovski and Gurzadyan (DG) (\cite{DG}) did an in-depth refinement while comparing these models to the observational data; in other words, they reconsidered the concept of one 
of the key parameters in the GX formula, the maximum distance $L_{max}$ assigned to the quantum fluctuations. By setting this distance to co-moving horizon, they got a 
parameter-free integral equation, which they solved iteratively. In this paper we  switch the DG equation to a differential one, avoiding the arising of an 
integration constant, and obtain a value of dark energy density again without any other assumptions except on the empirical values of the matter density and the 
curvature of the Universe; we also  derive the Hubble diagram for available SN and GRB data samples.

The paper is organized as follows. In the first part we derive the main formulae and get a numerical solution for parameter $y(z)$. Then we construct Hubble diagrams 
for SN and GRB samples and compare Hubble parameter graphs for the lambda cold dark matter ($\Lambda$CDM) model and the GX-DG model. While the observational data seem 
to support the EOS parameter $w(z)\approx-1$ for dark energy (\cite{Astier,Riess2,Riess3}), we derive an equation for $w(z)$ and see that it slightly varies with redshift. 
We  also consider some further cosmological insights on the considered dark energy scalings. 

\section{Dark energy versus  luminosity distance}

In their original work \citet{GX,GX1}  proposed  considering, in the dark energy only, the contribution of $l=0$ spherical modes of vacuum fluctuations. As was well known, 
accounting for all vacuum fluctuations leads to about $10^{123}$ quantitative discrepancies of the theoretical density with the empirical dark energy value. Thus, they derived 
a simple equation for dark energy density,
\begin{equation}
\Omega_{\Lambda}=\frac{\pi^2}{3}\left(\frac{D_H}{L_{max}}\right)^2,
\label{omega_lambda}
\end{equation}
where ${D}_{H}=c/H_{0}$ is the Hubble horizon scale, i.e., the distance that light would travel during Hubble time $1/H_0$.

The key issue with Eq.(\ref{omega_lambda}) is what to choose for the maximum distance  $L_{max}$ associated with the fluctuations. In some papers 
(\cite{Ver06,Ver06a,Ver06b,Ver07,Khach07a,Khach07}) the scale factor is used as a maximum distance and that assumption gave  $\Omega_{\Lambda}\approx3.29$ for dark energy density, 
while the observed value is $\approx0.7$). 

\citet{DG} suggested another solution for the choice of the maximum distance $L_{max}$, namely using the distance to the horizon measured in co-moving units
\begin{eqnarray}
D_{\infty}(z)&=&D_{H}\int_{z}^{\infty}\frac{1}{h(\acute{z})}d\acute{z}\\ \nonumber
h(z)&=&\sum_{i}\Omega_i^{(0)}(1+z)^{3(1+\omega_i)},
\label{D_inf}
\end{eqnarray}
where $D_{\infty}$ and $h(z)=H(z)/H_0$ are the distance to the horizon and a dimensionless Hubble parameter, respectively. As a result they obtained only a 20\% difference 
between the theoretical and observed values of the dark energy density. This integral equation can be rewritten in the form of differential equation for $y(z)$ parameter 
\begin{eqnarray}
\frac{dy}{dz}&=&-\frac{1}{\sqrt{{\Omega_m}(1+z)^3 + {\Omega_r}(1+z)^4 + \frac{\pi^2}{3y^2}}},\\ \nonumber
y'(0)&=&-1,\\ \nonumber
y(z)&=&\frac{D_{\infty}(z)}{D_H}.
\label{dydt}
\end{eqnarray}
This equation assumes that in the early Universe $z \rightarrow \infty$ the Hubble parameter $h(z) \rightarrow \infty$.  

The initial condition for $y'$ follows from the Friedmann equations and can be rewritten as $\sum{\Omega_i^{(0)}}=1$. Here we consider only flat $k=0$, $\Omega_{k}=0$ 
case, but the equation is true for any $k$. 

\begin{figure}[htb]
  \includegraphics[width=0.48\textwidth]{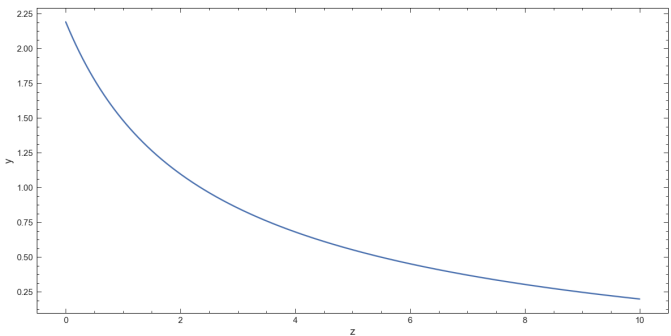}
  \caption{Numerical solution of Eq.(\ref{dydt}) for redshift $z$ up to 10. Model parameters are set to $\Omega_m=0.315, \Omega_r=1.08622\cdot10^{-5}$.}
        \label{y_z}
\end{figure}

Eq.(\ref{dydt}) cannot be solved analytically and the numerical solution is depicted in Fig.(\ref{y_z}).

To construct the Hubble diagram the luminosity distance and the distance moduli are needed. They can be expressed by $y(z)$ 
in a simple form
\begin{eqnarray}
d_L(z)&=&D_{H}(1+z)(y(0)-y(z)),\\ \nonumber
d_M(z)&=&5\log_{10}(\frac{d_L(z)}{10\,pc})
\label{d_lm}
\end{eqnarray}

\section{Hubble diagram}

\begin{figure}[htb]
        \includegraphics[width=0.48\textwidth]{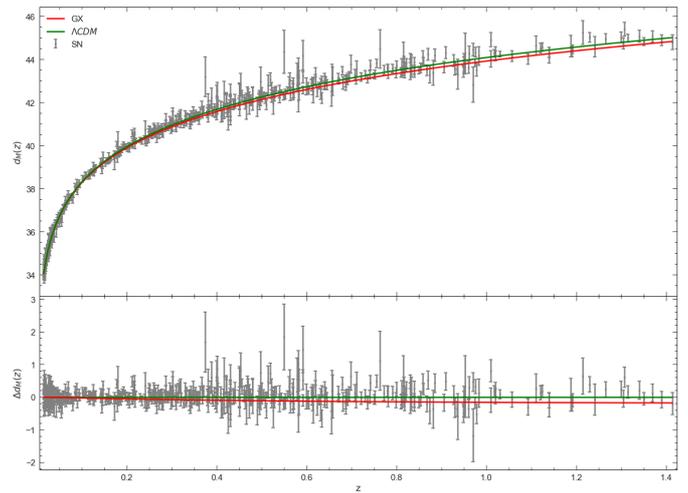}
        \caption{Hubble diagram for 580 SNs of the Union2.1 SN Ia sample (\cite{Amanullah,Suzuki}). The red solid line represents the GX-DG cosmological model and the green 
                 solid line the $\Lambda{CDM}$ model. The Hubble constant value is taken as $H_0=70\,km\,s^{-1}\,{Mpc}^{-1}$.}
        \label{hd_sn}
\end{figure}

In this section we construct the Hubble diagram for the Union2.1 sample of SN Ia (\cite{Amanullah,Suzuki}). We used a standard parameter set for the numerical solution of 
$y(z)$ (see Fig.(\ref{y_z})). It shows that the standard model with a cosmological constant predicts values that are a bit higher for distance moduli of SNs (Fig.(\ref{hd_sn})). 
The  $\chi^2$ values for $\Lambda$CDM model are somewhat smaller than for the DG-GX model, $\chi^2=1.1$ and $\chi^2=1.29$, respectively.

We also fit the models with GRB data from \cite{Demianski}, which were calibrated independently using Amati relation and can be used with any dark energy model. The Amati relation (\cite{Amati2006}) refers to an empirical relation between isotropic equivalent radiated energy $E_{iso}$ and the observed photon energy of the peak spectral flux$E_{p,i}$: 

\begin{equation}
\log\left(\frac{E_{iso}}{1\,erg}\right)=b+a\log\left[\frac{E_{p,i}}{300\,KeV}\right].
\label{amati}
\end{equation}

The Amati relation is known to be not applicable for any GRBs with measured parameter pairs $E_{iso},E_{p,i}$ (see \cite{Qin2013,Lin2015}); moreover, the regression parameters $a,b$ 
slightly differ when obtained by different authors. Although the Amati relation was nevertheless concluded to be the preferable to other empirical relations for GRBs while constructing 
the Hubble diagram \cite{Wang2020},  more detailed work is indeed needed in the future in the comparative analysis of various empirical relations and their possibly combined applications.

\begin{figure}[htb]
  \includegraphics[width=0.48\textwidth]{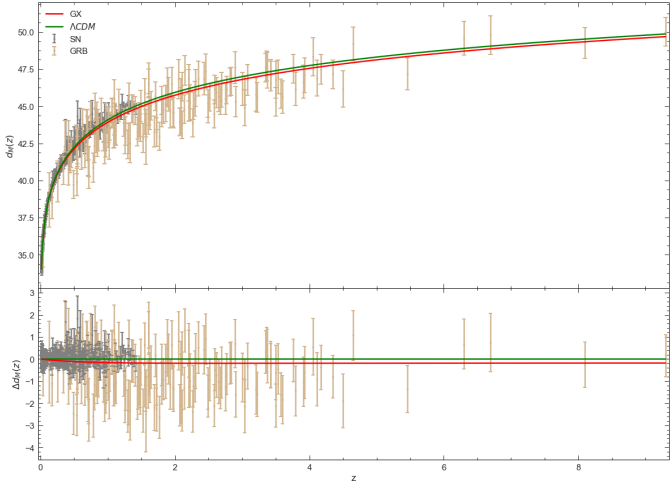}
        \caption{Hubble diagram for 580 SNs of the Union2.1 SN Ia sample and 162 calibrated GRBs from \cite{Demianski}. The red line 
                 represents the GX-DG cosmological model and the green line the $\Lambda{CDM}$ model.}
        \label{hd_sn_grb}
\end{figure}

For GRBs we see that the GX-DG model fits better: for the $\Lambda$CDM model we have $\chi^2=1.41$ and for the GX-DG model $\chi^2=1.21$. This can be seen even in Fig.(\ref{hd_sn_grb}), 
where the GRB distance moduli values are smaller and therefore the GX-DG model fits better. To trace the statistical properties of the GRB data we fitted and then calculated $\chi^2$, 
for example  for the GRB data from \cite{Liu2015} as well; we obtained lower values for $\chi^2$ ($0.4271,\,0.4319$) for the GX-DG and $\Lambda$CDM models, respectively.

\begin{figure}[htb]
        \includegraphics[width=0.48\textwidth]{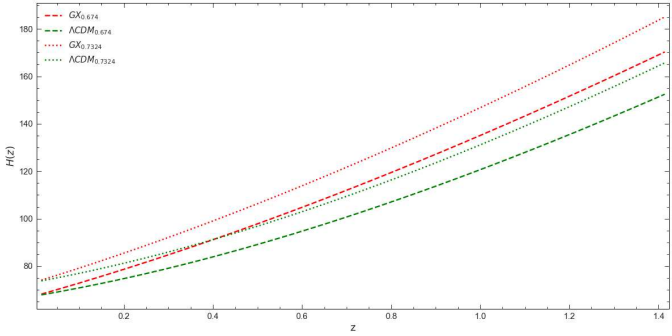}
        \caption{Hubble parameter $H(z)$ dependance on the redshift $z$. The red lines represent GX-DG cosmological model and the green lines refer to the $\Lambda{CDM}$ model 
                 (dashed $H_0=73.24\,km\,s^{-1}\,{Mpc}^{-1}$, dotted $H_0=67.4\,km\,s^{-1}\,{Mpc}^{-1}$.}
        \label{H_z}
\end{figure}

Then we plot in Fig.(\ref{H_z}) the Hubble parameter dependence on the redshift for different values $H_0=67.4,73.24\,km\,s^{-1}\,{Mpc}^{-1}$ for the $\Lambda$CDM and the 
GX-DG model; we chose the Hubble parameter because the \textit{Planck} satellite data on the Cosmic Microwave Background (CMB) \cite{Planck} provide the Hubble constant 
value $H_0=67.4\pm0.5\,km\,s^{-1}\,{Mpc}^{-1}$, while the Milky Way Cepheid survey (\cite{Riess4,AR}) suggests $H_0=73.24\pm1.62\,km\,s^{-1}\,{Mpc}^{-1}$. We note 
that the graph with the GX-DG model for the Hubble constant value $H_0=67.4\,km\,s^{-1}\,{Mpc}^{-1}$ crosses at around $z=0.4018$ the graph of the $\Lambda$CDM model 
with Hubble constant value $H_0=73.24\,km\,s^{-1}\,{Mpc}^{-1}$;  i.e., at redshifts lower than $z=0.4018$ the Hubble parameter for GX-DG model approximate $\Lambda$CDM 
model's graph with $H_0=67.4\,km\,s^{-1}\,{Mpc}^{-1}$ and at higher redshifts tending to that of $H_0=73.24\,km\,s^{-1}\,{Mpc}^{-1}$.

The interpretation of this result, and whether it is related to the Hubble parameter issues in \cite{VTR}, is not clear. We can predict, to be confirmed or ruled out at future 
observations, that at $z=0.4018$ either the Hubble parameter or the parameter $w$, or some other related parameter undergoes a change in properties.

\section{Equation of state parameter}

In the standard $\Lambda$CDM model dark energy is adopted as the cosmological constant. Therefore, it is important to have an equation for the EOS parameter $w(z)$    
dependence on redshift for the GX-DG model; for $\Lambda$CDM $w(z)\approx-1$. From Eq.(\ref{dydt}) we can easily derive a formula for the EOS parameter
\begin{equation}
w(z)=\frac{2(1+z)}{3h(z)y(z)} - 1.
\label{wz_eq}
\end{equation}

\begin{figure}[htb]
        \includegraphics[width=0.48\textwidth]{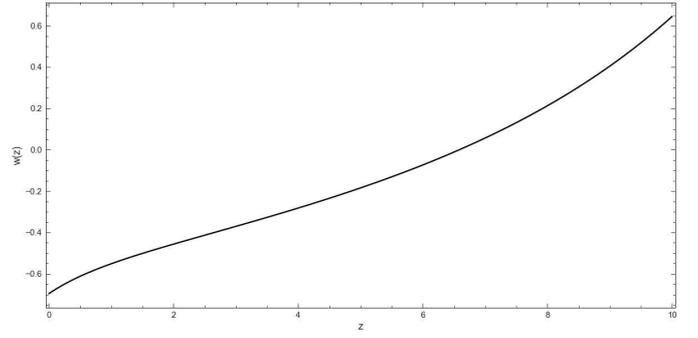}
        \caption{Effective EOS parameter $w(z)$ as a function of redshift $z$.}
        \label{w_z}
\end{figure}

In Fig.(\ref{w_z}) we can see the dependance of the EOS parameter on the redshift. It is negative in a wide range of redshifts, but at around $z\approx6$ it turns to zero. 

\section{Information evolution}

In this section we link the GX-DG model  to the ``information'' description of the cosmological evolution (\cite{GS1}). Thus, the vacuum matter density is 

\begin{equation}
\rho \propto (L_{pl})^{-2}(L_{max})^{-2},
\label{rho}
\end{equation}
where $L_{pl}$ is the Planck length and  $L_{max}$ is the upper bound of the vacuum modes, which is considered the cosmological horizon in co-moving coordinates \citep{DG}.

Within the framework of the information evolution (IE), it is assumed that the evolution of the Universe can be described according to the increase of a 
discretized quantity, the information. This concept is obtained based on the notion of Bekenstein Bound (\cite{Bekenstein}), which introduces an upper bound of information
\begin{equation}
I_{BB} \leq \frac{2 \pi R E}{\hbar c \ln{2}},
\label{BB}
\end{equation}
where $R$ and $E$ are the characteristic radius and energy of a system under the consideration. Thus, the evolution of the Universe ($T$) is discretized as follows:
\begin{equation}
T= \left\{1, 2 , ...j, ..., 3\pi I\right\}, \quad I=\frac{c^3}{\Lambda G \hbar}.
\label{EvUI} 
\end{equation}
Accordingly, at each stage ($j$) of IE, the Universe is denoted by the characteristic mass $M_j$ and radius $R_j$
\begin{equation}
I_{BB}(M_j, R_j) = I_{BB}(\frac{\pi c^8}{2 j \hbar G^2 H^3_j}, \frac{c}{H_j}) \quad j=1,..., \frac{3 \pi c^3}{\hbar G \Lambda}
\label{Area}
,\end{equation}
where $\frac{c}{H_j}$ is the Hubble horizon at that stage. 
On the one hand, $I_{BB}$ is inversely proportional to the total density of the Universe and Eq.(\ref{EvUI}) can be written as 

\begin{equation}
T= \left\{\rho _1 = \frac{3}{8}\frac{c^5}{\hbar G^2}, ...\rho _j = \frac{1}{j}\rho_{1}..., 
                    \rho _{3\pi I} = \frac{1}{3\pi I}\rho_{1} = \frac{\Lambda c^2}{8 \pi G}\right\}.
\label{EvU} 
\end{equation}
On the other hand, the evolution of the Universe can be described based on the area of the cosmological horizon at each stage $A_j$, i.e.,
\begin{equation}
T= \left\{A_1 = 4l_p^2, ...A_j = j A_1, ..., A_{3\pi I} = (3\pi I) A_1 = \frac{12 \pi}{\Lambda} \right\},
\label{EvArea}
\end{equation}
so that the information content of Universe increases until the area of the Hubble horizon reaches $\frac{12 \pi}{\Lambda}$, the de Sitter cosmological horizon. 
Consequently, at this stage the density becomes $\frac{\Lambda c^2}{8 \pi G}$. Thus, in the context of IE
\begin{equation}
\text{Information} \propto \text{Area} \propto (\text{density})^{-1}.
\label{IdA}
\end{equation}
Then, the ``maximum number of relevant radial modes'' in the GX model is defined as
\begin{equation}
\rho \propto N_{max}(N_{max} + 1), \quad N_{max} = \frac{L_{max}}{L_{pl}}.
\label{MaxN}
\end{equation}
Comparing these relations with the notion of area in IE leads us to the following degrees of freedom
\begin{equation}
I_{j} = \frac{A_{j}}{A_{pl}} = \frac{4 \pi (\frac {c}{H_{j}})^2 }{4 \frac{G \hbar}{c^3}}.
\label{IN}
\end{equation}

Thus, by comparing Eq.(\ref{rho}) with the notions of ``area of cosmological horizon'' and ``matter density'' we can conclude that the results of GX-DG and IE models, although 
assuming entirely different approaches, are similar. It should be noted that this similarity exists only if $\rho$ in Eq.(\ref{rho}) corresponds to the total density of the 
Universe (according to Eq.(\ref{EvU})).

Obviously, the two approaches deal with certain global scaling parameters and they should be reduced to standard Friedmann-Lemaitre-Robertson-Walker universe with the continuous geometrical structure. 

\section{Conclusions}

We reduced the \citet{DG} integral equation to a differential equation to describe   dark energy with observable parameters. Solving that differential equation we obtained the 
exact value of dark energy density. This allowed us to construct the Hubble diagram for SN and GRB samples and to show that the model estimates the observed distances up to 
$z\simeq 9$ at least as well as the standard $\Lambda$CDM model with a cosmological constant. The dark energy EOS parameter predicts a slight variation with redshift which  
is still not observable, however. 

The behavior of the GX-DG model and comparison with observational data shows that if there are no unknown experimental systematics (\cite{Niedermann,Efstathiou1,Efstathiou2}), the GX-DG 
model can be  adopted along with the standard $\Lambda$CDM model. The revealed feature of the diagrams, namely, the crossing of the curves of both models at $z=0.4018$ is not obvious for 
direct interpretation, suggesting possible different behaviors of parameters at redshifts $z<0.4018$ and $z>0.4018$, as a test to be solved by means of future observations. Finally, regarding 
the DG integral equation we note that the first-order differential equation for dark energy density Eq.(\ref{dydt}) has one integration invariant that can be related to the invariants of the general equations of GX models (\cite{Khach07}).

\section{Acknowledgments}
We are thankful to the referee for very useful comments and suggestions that helped to improve the paper.

\end{document}